\documentclass[12pt]{article}
\usepackage{amsmath,amssymb}
 \usepackage{graphicx}

\begin{document}

\begin{center}

{\Large\bf Eigenfrequencies of nodeless elastic vibrations locked in the crust of quaking
 neutron star}
 \vspace{1.cm}

S.I. BASTRUKOV\footnote{Joint Institute for
 Nuclear Research,
141980 Dubna, Russia. E-mail: bast@jinr.ru},\, H.-K. CHANG,\, G.-T. CHEN

Department of Physics and  Institute of Astronomy,
  National Tsing Hua University, Hsinchu, 30013, Taiwan\\
 bast@phys.nthu.edu.tw; hkchang@phys.nthu.edu.tw; d923317@oz.nthu.edu.tw

I. V. MOLODTSOVA

Joint Institute for
 Nuclear Research,
141980 Dubna, Russia, e-mail molod@theor.jinr.ru

\end{center}

\begin{abstract}
The Newtonian solid-mechanical theory of non-compressional spheroidal and torsional nodeless elastic vibrations
 in the homogenous crust model of a quaking neutron star is developed and applied to
 the modal classification of the quasi-periodic oscillations (QPOs) of X-ray luminosity in the aftermath
 of giant flares in SGR 1806-20 and SGR 1900+14. Particular attention is given to the low-frequency QPOs in the data for SGR 1806-20 whose physical origin has been called into question.
 Our calculations
 suggest that unspecified QPOs are due to nodeless dipole torsional and dipole spheroidal elastic shear
 vibrations.
\end{abstract}



 Published in {\sf Mod. Phys. Lett. A, Vol. 23, No. 7 (2008) pp. 477-486.}

\section{Introduction}
 This study was undertaken as a part of current extensive theoretical investigations of asteroseismology of a neutron
 star\cite{P-05,GSA-06,SA-07,Lee-07,Lev-07,Bast-07} (see also references therein) that have been
 boosted by recent discovery of quasi-periodic oscillations (QPOs) of X-ray luminosity
 in the decaying flares of two magnetars\cite{Isr-05,SW-06}, SGR 1806-20 and SGR 1900+14, with concomitant
 suggestion to interpret this variability as caused by quake induced differentially rotational, torsional, oscillations.  Following this suggestion the focus of  most theoretical works is on computing the frequency spectra of odd-parity torsional mode of shear vibrations and less attention is paid to the even parity spheroidal
 elastic mode. However, from the viewpoint of modern global seismology\cite{LW-95,AR-03}, the spheroidal vibrational mode in a solid star and planet
 has the same physical significance as the toroidal one in the sense
 that these two fundamental modes owe their existence to one and the same restoring
 force\cite{McD-88,Bast-99,Bast-IJ-07}. In this light there is a possibility
 that, by not considering both these modes on an equal footing, we may
 miss discovering certain essential novelties which are consequences of solid mechanical laws governing
 seismic vibrations of superdense matter of neutron stars.
 Adhering to this attitude and continuing our current investigations\cite{Bast-IJ-07},
 we derive here spectral equations for the frequency of both spheroidal and torsional elastic nodeless vibrations
 in the solid crust of quaking neutron star and examine what conclusions can be drawn regarding
 low-frequency QPOs whose physical nature still remain unclear.

 \section{Frequency of nodeless spheroidal and torsional elastic shear vibrations in homogeneous crust}

 In this paper we follow the line of argument of the standard two-component, core-crust, model of quaking
 neutron star\cite{FLE-00} in which crustal metal-like material (composed of nuclei dispersed in the sea of
 relativistic electrons) is treated as
 a highly robust to compressional distortions elastic material continuum of a uniform density
 $\rho$ characterized by constant value of shear modulus $\mu$. In this model it is presumed that
 the quake-induced non-compressional seismic vibrations driven by bulk force of pure
 shear elastic deformations (which are not accompanied by fluctuations in density $\delta
 \rho=-\rho\,\nabla_k u_k=0$) can be adequately modeled by equation
 of Newtonian, non-relativistic, solid mechanics
 \begin{eqnarray}
 \label{e2.1}
&& \rho{\ddot u}_i=\nabla_k\sigma_{ik},\quad \sigma_{ik}=2\,\mu\,
u_{ik},\quad u_{ik}=\frac{1}{2}[\nabla_i u_k+\nabla_k u_i],\quad
u_{kk}=\nabla_k u_k=0.
 \end{eqnarray}
 From now on $u_i({\bf r},t)$ stands for the field of material displacements in the crust
 of the depth $\Delta R=R-R_c$ with $R$ and $R_c$ being radii of star and core, respectively.
 The linear relation between tensors of shear elastic stresses $\sigma_{ik}$ and
 shear deformations or strains $u_{ik}$ is the
 Hooke's law of elastic (reversal) shear deformations.

 In what follows we focus on poorly investigated regime of nodeless shear vibrations
 in which the fields of oscillating
 material displacement subject to the vector Laplace equation\cite{Bast-IJ-07}
  \begin{eqnarray}
  \label{e2.2}
 && \nabla^2 {\bf u}({\bf r},t)=0.
 \end{eqnarray}
 This last equation can be thought of as the long wavelength limit of vector Helmholtz
 equation describing standing-wave regime of vibrations\cite{Bast-07,Bast-IJ-07}.
 In these latter works it has been shown that the eigenfrequency problem of
 global nodeless vibrations of a solid star can be unambiguously
 solved with aid of the Rayleigh's energy method. In the present paper this method is extended to
 the case of spheroidal nodeless vibrations which are considered in one line with torsional ones.

 The stating point of the energy variational method is the integral
 equation of the energy balance
\begin{eqnarray}
 \label{e2.3}
  \frac{\partial }{\partial t}\int \frac{\rho {\dot u}^2}{2}\,d{\cal
  V} = -\int \sigma _{ik}{\dot u}_{ik}\,d{\cal V}=-2\int \mu\, u_{ik}{\dot u}_{ik}d{\cal
  V}
   \end{eqnarray}
which is obtained by scalar multiplication of equation of solid
mechanics, (\ref{e2.1}), with $u_i$ and integration over the volume
of seismogenic layer. The field ${\bf u}({\bf r},t)$ can be
conveniently represented in the following separable form
\begin{eqnarray}
 \label{e2.4}
 {\bf u}({\bf r},t)={\bf a}({\bf r})\,\alpha(t)
 \end{eqnarray}
with the field of instantaneous (time-independent) displacements
${\bf a}({\bf r})$ obeying, as follows from (\ref{e2.2}), to
equations
\begin{eqnarray}
  \label{e2.5}
 && \nabla^2 {\bf a}({\bf r})=0,\quad\quad \nabla \cdot {\bf a}({\bf r})=0
 \end{eqnarray}
and $\alpha(t)$ stands for the temporal amplitude of vibrations.
Inserting (\ref{e2.4}) in (\ref{e2.3}) we arrive at equation for
${\alpha}(t)$ having the form of standard equation of normal
oscillations
\begin{eqnarray}
 \label{e2.6}
 && \frac{dE}{dt}=0,\quad E=\frac{M{\dot\alpha}^2}{2}+\frac{K{\alpha}^2}{2}
 \quad\to\quad {\ddot\alpha}+\omega^2\alpha=0,\quad\quad \omega^2=\frac{K}{M},\\
 \label{e2.7}
 && M=\int \rho\, a_i\,a_i\,d{\cal V},\quad\quad
 K=2\int \mu\, a_{ik}\,a_{ik}\,d{\cal V}\quad  \quad a_{ik}=\frac{1}{2}[\nabla_i a_k + \nabla_k
 a_i].
 \end{eqnarray}
 The solenoidal fields of instantaneous material
 displacements in two fundamental modes of nodeless vibrations --
 the spheroidal (normally abbreviated as
 $_0s_\ell$) and the toroidal (abbreviated
 as $_0t_\ell$), are determined by two fundamental (orthogonal and different in
 parity) solutions to the vector Laplace equation which are uniquely defined by the general solution to the
 scalar Laplace equation $\nabla^2\chi({\bf r})=0$. In spherical coordinates with fixed polar axis, the solution
 of (\ref{e2.5}) corresponding to nodeless spheroidal vibrations, ${\bf a}_s$, is given by
 the even parity poloidal (polar) vector field
 and instantaneous displacements in the torsional mode, ${\bf a}_t$, are  described by
 the odd parity toroidal (axial) vector field:
 \begin{eqnarray}
 \label{e2.8}
&& {\bf a}_s=\nabla \times \nabla\times\,({\bf
 r}\,\chi),\quad\quad {\bf a}_t=\nabla \times \, ({\bf r}\chi),\\
&&
 \chi({\bf r})=f_\ell({\bf r})P_\ell(\cos\theta),\quad\quad
 f_\ell({\bf r})=[{\cal A}_\ell\,r^\ell+{\cal
  B}_\ell\,r^{-(\ell+1)}].
 \end{eqnarray}
 Henceforth $P_\ell(\cos\theta)$ stands for the Legendre polynomial of multipole
 degree $\ell$ and ${\cal A}_{\ell}$ and ${\cal B}_{\ell}$
 are the arbitrary constants to be eliminated from boundary conditions on the core-crust interface and
 on the star surface. Thus, in order to obtain the frequency
 spectra of both spheroidal and torsional nodeless vibrations one need
 to specify first these constants and then to compute integrals for integral parameters
 of vibrations, that is, the inertia $M$ and the stiffness $K$. Some
 useful mathematical details of such calculations can be found in\cite{Bast-IJ-07}.

 \subsection{Spheroidal mode}
  The poloidal field of  nodeless instantaneous displacement ${\bf a}_s$ in s-mode
 is irrotational: $\nabla\times {\bf a}_s=0$.
 To specify ${\cal A}_{\ell}$ and ${\cal B}_{\ell}$
 we adopt on the core-crust interface, $r=R_c$, the condition
 of impenetrability of seismic perturbation in the core. On the star edge, $r=R$, we impose the condition that
 the radial velocity of material displacements equals the rate of spheroidal distortions of the star
 surface\cite{Lamb}
  \begin{eqnarray}
  \label{e3.1}
  u_r\vert_{r=R_c}=0,\quad\quad {\dot u}_r\vert_{r=R}={\dot R}(t),\quad\quad
  R(t)=R[1+\alpha(t)\,P_\ell(\cos\theta)].
 \end{eqnarray}
 The solution of resultant algebraic equations leads to following
 values of arbitrary constants
\begin{eqnarray}
 \label{e3.2}
  {\cal A}_\ell=\frac{{\cal N}_\ell}{\ell(\ell+1)},\quad
  {\cal B}_\ell=-\frac{{\cal N}_\ell}{\ell(\ell+1)}\,R_c^{2\ell+1},\quad
  {\cal N}_\ell=\frac{R^{\ell+3}}{R^{2\ell+1}-R_c^{2\ell+1}}.
  \label{e2.10}
  \end{eqnarray}
  Tedious but simple calculation of integrals for inertia $M$ and stiffness $K$, given by (\ref{e2.7}),
  with poloidal field ${\bf a}_s$  yields
 \begin{eqnarray}
  \label{e3.3}
  && M_s(\ell,\lambda)
=\frac{4\pi R^5\rho}{\ell(2\ell+1)(1-\lambda^{2\ell+1})}
\left[1+\frac{\ell}{(\ell+1)}\lambda^{2\ell+1}
\right],\\
 \label{e3.4}
&& K_s(\ell,\lambda) =8\pi R^3 \mu
\frac{(\ell-1)(1-\lambda^{2\ell-1})}{\ell(1-\lambda^{2\ell+1})^2}\left[
 1 +
 \frac{\ell(\ell+2)}{\ell^2-1}\,\frac{\lambda^{2\ell-1}(1-\lambda^{2\ell+3})}{(1-\lambda^{2\ell-1})}\right],\\
 \label{e3.5}
 && \lambda=\frac{R_c}{R}=1-h\quad\quad h=\frac{\Delta R}{R}.
 \end{eqnarray}
 The fractional frequency of nodeless spheroidal irrotational shear vibrations
 as a function of multipole degree $\ell$
 is given by
 \begin{eqnarray}
 \label{e3.6}
&&\frac{\omega^2_s(\ell)}{\omega_0^2}=\frac{2(2\ell+1)}{(1-\lambda^{2\ell+1})}
\left[\frac{(\ell^2-1)(1-\lambda^{2\ell-1})+\ell(\ell+2)\lambda^{2\ell-1}(1-\lambda^{2\ell+3})}
 {(\ell+1)+\ell\lambda^{2\lambda+1}}\right],\\
  \label{e3.7}
&& \omega_0=\frac{c_t}{R},\quad c_t=\sqrt{\frac{\mu}{\rho}},\quad
[{\omega_0}={2\pi}\nu_0,\quad {\omega_s(\ell)}={2\pi}\nu(_0s_\ell)].
\end{eqnarray}
 It worth noting that in the limit of zero-size radius of the core,
 $\lambda=(R_c/R)\to 0$, when entire volume of the star sets in
 vibrations, we regain the early obtained spectral formula for global
 nodeless spheroidal nodeless shear vibrations\cite{Bast-IJ-07}
 $\nu(_0s_\ell)=\nu_0\,[2(2\ell+1)(\ell-1)]^{1/2}$ showing that the
 lowest overtone of the global
 nodeless spheroidal oscillations in the entire volume of the star
 is of quadrupole degree, $\ell=2$ (see also\cite{Bast-02}). In the meantime,
 the lowest overtone of spheroidal vibrations trapped in the crust is of the dipole
 degree, $\ell=1$. This suggests that the dipole overtone can be considered
 as a signature of spheroidal vibrations locked in
 the crust. It is remarkable that dipole vibration can be thought of as, so called,
 Goldstone's soft mode whose most conspicuous feature is that
 the frequency as a function of intrinsic parameter $\lambda$ of oscillating
 system $\omega_s(\ell=1,\lambda)\to 0$, when $\lambda\to 0$.
 In the model under consideration this parameter is given by
 $\lambda=(R_c/R)$. The limit $\lambda=0$ belongs to
 translation displacement of the center-of-mass of the star, not a vibration; this is clearly seen
 from the equation for energy (Hamiltonian) of harmonic oscillations
 (\ref{e2.6}).

  \begin{figure}
 \centering{\includegraphics[width=8.cm]{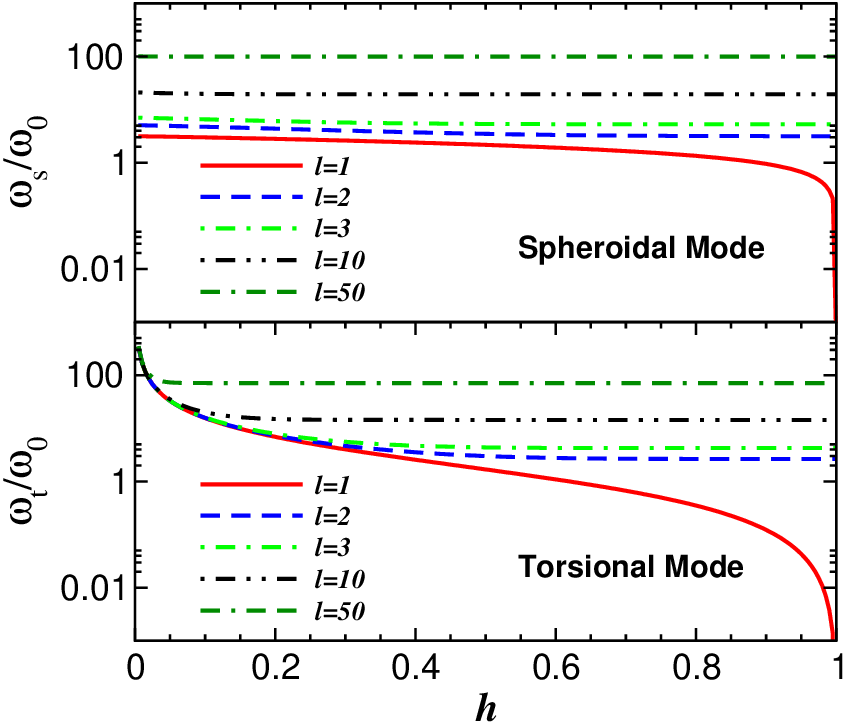}}
 \caption{
 Fractional frequency of nodeless spheroidal and torsional elastic oscillations as a function of fractional depth
 of seismogenic layer.}
\end{figure}

 In the upper panel of Fig.1 we plot the fractional frequency $\omega_s(\ell)/\omega_0$
 as a function of $h=\Delta R/R$.
 In the next section we show how the input
 parameters of obtained spectral equation (\ref{e3.6}), namely, the natural unit of
 frequency $\nu_0$ and the depth $h$ of
 seismogenic layer can be extracted from the data on QPOs for SGS.

\subsection{Torsion mode}
  For the torsional oscillations locked in the crust, the constants ${\cal A}_\ell$ and
 ${\cal B}_\ell$ are eliminated from the following boundary conditions
  \begin{eqnarray}
   \label{e4.1}
 && u_\phi\vert_{r=R_c}=0,\quad u_{\phi}\vert_{r=R}=[\mbox{\boldmath $\phi$}_R\times {\bf R}]_\phi, \\
  \label{e4.2}
 && \mbox{\boldmath $\phi$}_R=\alpha(t)\nabla_{\hat{\bf n}} P_\ell(\zeta),\quad\quad
 \nabla_{\hat{\bf n}}=\left(0,\frac{\partial }{\partial
 \theta},\frac{1}{\sin\theta}\frac{\partial }
 {\partial \phi}\right).
 \end{eqnarray}
 First is the no-slip condition on the core-crust interface, $r=R_c$,
 implying that the amplitude of differentially rotational oscillations
 is gradually decreasing from the surface to the core.
 The boundary condition on the star surface, $r=R$,
 is dictated by symmetry of the general toroidal solution of the vector Laplace equation.
 The support of this last boundary condition lends further considerations showing that
 it leads to correct expression for the moment of inertia of a rigidly rotating star.
 The resultant algebraic equations steaming from above boundary conditions
 lead to
  \begin{eqnarray}
  \label{e4.3}
 {\cal A}_\ell={\cal N}_\ell,\quad {\cal B}_{\ell}=-{\cal
 N}_\ell\,R_c^{2\ell+1},\quad\quad {\cal
 N}_\ell=\frac{R^{\ell+2}}{R^{2\ell+1}-R_c^{2\ell+1}}.
 \end{eqnarray}
 Tedious calculation of integrals for $M_t$ and $K_t$ leads to
 \begin{eqnarray}
  \label{e4.4}
 && M_t=\frac{4\pi\ell(\ell+1)}{(2\ell+1)(2\ell+3)}\frac{\rho R^5}{(1-\lambda^{2\ell+1})^2}\times\\
  \nonumber
 &&\left[1- (2\ell+3)\lambda^{2\ell+1}+
\frac{ (2\ell+1)^2}{2\ell-1}
\lambda^{2\ell+3}-\frac{2\ell+3}{2\ell-1}\lambda^{2(2\ell+1)}\right],\\[0.5cm]
 \label{e4.5}
 && K_t=\frac{4\pi\ell(\ell^2-1)}{2\ell+1}\,\frac{\mu R^{3}}{(1-\lambda^{2\ell+1})}
 \left[1+\frac{(\ell+2)}{(\ell-1)}\lambda^{2\ell+1}\right]\\
 && \lambda=\frac{R_c}{R}=1-h\quad h=\frac{\Delta R}{R}.
\end{eqnarray}
In the limit of zero-size radius of the core, $\lambda=(R_c/R)\to
 0$, corresponding to torsional oscillations in the entire volume of
the star  we regain the early obtained spectral formula for the
global nodeless torsional elastic vibrations
$\nu(_0t_\ell)=\nu_0\,[(2\ell+3)(2\ell-1)]^{1/2}$ showing that in
case of global torsional oscillations the lowest overtone is of
quadrupole degree\cite{Bast-07,Bast-IJ-07,Bast-02}.
However, this is not the case when we consider torsional nodeless
oscillations locked in the seismogening layer of finite depth
$\Delta R=R-R_c$. For $\ell=1$, equations (\ref{e4.4}) and
(\ref{e4.5}) are reduced to
\begin{eqnarray}
  \label{e4.6}
 && M_t(\ell=1,\lambda)=\frac{8\pi\,\rho\,R^5}{15(1-\lambda^3)^2}\left[1-5\lambda^3+9\lambda^5-5\lambda^6\right],\\
 \label{e4.7}
 && K_t(\ell,\lambda)=8\pi\,\mu \,R^3\,\frac{\lambda^3}{(1-\lambda^3)},\\
  \label{e4.8}
 && \omega_t^2(\ell=1,\lambda)=\omega_0^2\,
 \frac{15\lambda^3(1-\lambda^3)}{(1-\lambda)^3(1+3\lambda+6\lambda^2+5\lambda^3)}\quad  0\leq \lambda <  1.
\end{eqnarray}
 In the limit when the core radius tends to zero, the stiffness $K_t(\ell=1,\lambda=0)\to 0$ and the mass
 parameter getting the form of the moment of inertia of absolutely rigid
 solid star of mass ${\cal M}$ and radius $R$:
 $M_t(\ell=1,\lambda=0)=(2/5){\cal M}R^2$. This latter case corresponds to the rigid rotation.
 The above consideration again shows that the dipole overtone
 exhibits features of the Goldstone soft mode owing its emergence to the trapping of torsional shear
 oscillations in the peripheral crust of finite depth.

 The general spectral equation for the
 fractional frequency of nodeless torsional oscillations of arbitrary multipole degree
 $\ell$, computed with aid of equations (\ref{e4.4}) and
 (\ref{e4.5}), can be presented in the following analytic form
\begin{eqnarray}
 \label{e4.9}
 &&\frac{\omega_t^2(\ell)}{\omega_0^2}=\frac{\nu^2(_0t_\ell)}{\nu_0^2}=[(\ell+2)(\ell-1)]\,p_\ell(\nu_0,\lambda)\\
 \label{e4.10}
 && p_\ell(\nu_0,\lambda)=4\left[1-\frac{1}{2(\ell+2)}\right]\left[1+\frac{1}{2(\ell-1)}\right]
 (1-\lambda^{2\ell+1})\times\\
\nonumber
 &&\left\{1-
\frac{\ell-\lambda^{2\ell+1}[(\ell+2)+(2\ell-1)(2\ell+3)-(2\ell+1)^2\lambda^2+(2\ell+3)\lambda^{2\ell+1}]
}{\quad\,
(2\ell-1)-\lambda^{2\ell+1}[(2\ell-1)(2\ell+3)-(2\ell+1)^2\lambda^2+(2\ell+3){\lambda}^{2\ell+1}]
} \right\}.
\end{eqnarray}
The usefulness of such representation is extensively
discussed in\cite{Bast-07}. The fractional frequency as a
function of $h=\Delta R/R$ is pictured in down panel of Fig.1 which
show that the lowest overtone of torsional vibrations trapped in the
crust is of dipole degree and that dipole overtone of differentially
rotational vibrations of the crust against core possesses properties
of the Goldstone's soft mode.

\section{Application to SGR 1900+14 and SGR 1806-20}

 The obtained spectral formulae (15) and (26)-(27) describe the frequencies of both spheroidal and torsional
 nodeless oscillations
 as functions of the multipole degree $\ell$. The natural unit of frequency $\nu_0$ of shear elastic vibrations
 and the fractional depth $h$ of peripheral
 seismogenic layer are input parameters carrying information about material properties of
 neutron star matter (density, shear modulus) and geometrical sizes of star and seismoactive zone.
 Considering the observation data for SGR 1900+14 and SGR 1806-20, we demonstrate here how
 the obtained spectral equations can be used to eliminate some uncertainties
 in identification of QPOs.

\begin{figure}
\centering{\includegraphics[width=8.cm]{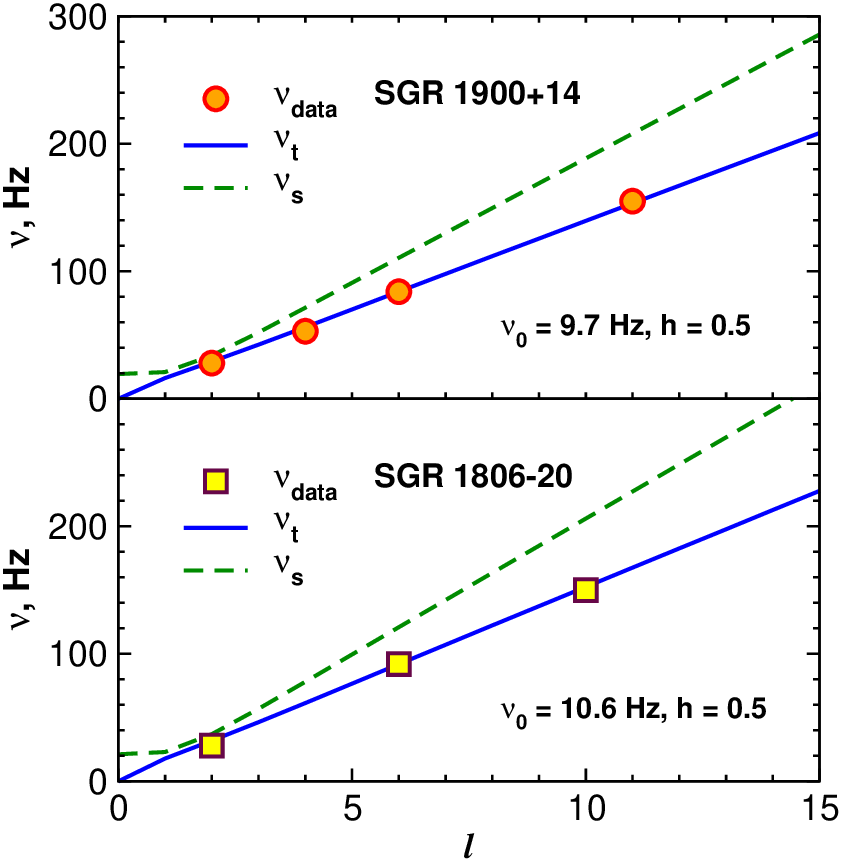}}
 \caption{
 Theoretical curves for the frequency of spheroidal (dashed) and torsional (solid) nodeless elastic
 oscillations computed with aid of spectral formulae for frequency of spheroidal, eq.(15), and
 torsional, eqs. (26)-(27),
 modes as functions of multipole degree in juxtaposition with data (symbols) on QPOs for SGR 1900+14
 and for SGR 1806-20.}
\end{figure}

 First, we examine the agreement of obtained  spectral formula (\ref{e4.9})-(\ref{e4.10}) for torsion mode
 with identification of the QPOs frequencies from interval $30\leq \nu\leq
 200$ Hz with frequencies of nodeless torsional vibrations of
 multipole degree $\ell$ from interval $2\leq \ell \leq 12$ suggested
 in\cite{SA-07}.
 In so doing we use the proposed in these latter works
 identification of QPOs in SGR 1900+14 data [namely, $\nu(_0t_2)=28$ Hz; $\nu(_0t_4)=53$; Hz
 $\nu(_0t_6)=84$ Hz, $\nu(_0t_{11})=155$ Hz borrowed from Table 1 of paper\cite{SA-07}] as reference points and vary parameters $\nu_0$ and $h$ entering
 our spectral formula (15)
 for torsional mode so as to attain the best fit of these points. The result of this
 procedure is shown in upper panel of Fig.2 by solid line.
 Then, making use of the fixed in the above manner parameters $\nu_0$ and $h$, we compute
 (with the aid of spectral formula (\ref{e3.6})) the frequency of spheroidal
 mode $\nu(_0s_\ell)$. The application of this procedure to  modal analysis of QPOs data for
 SGR 1806-20 is pictured in down panel of Fig.2. Based on proposed in the above mentioned paper identification
 of the following points $\nu(_0t_2)=30$ Hz;  $\nu(_0t_6)=92$ Hz and  $\nu(_0t_{10})=150$ Hz we
 extract parameters $\nu_0$ and $h$ entering in our spectral formulae for torsional
 mode, equations (\ref{e4.9})-(\ref{e4.10}).

 \begin{figure}
 \centering{\includegraphics[width=8.cm]{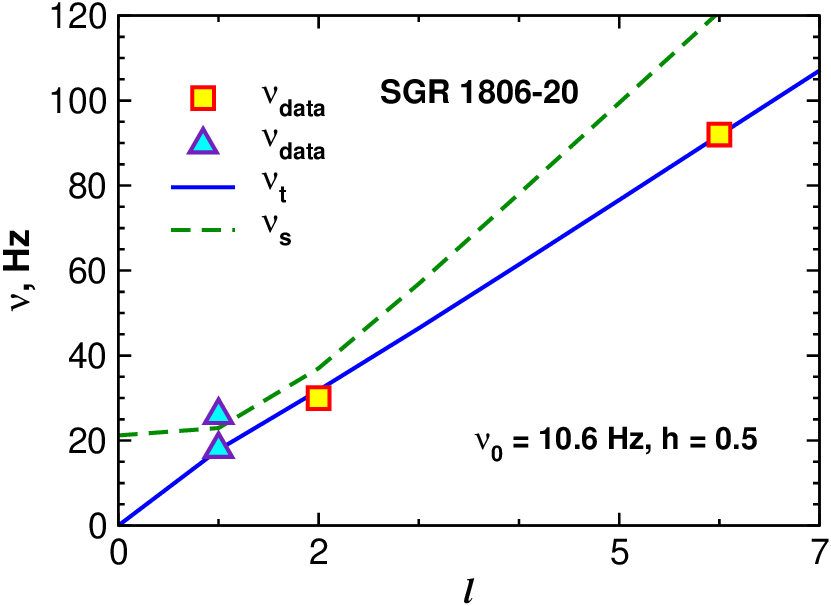}}
 \caption{
 Theoretical predictions for the frequency of spheroidal
 (dashed) and torsional  (solid) nodeless elastic
 oscillations as a function of multipole degree $\ell$ in juxtaposition with data (symbols) on QPOs
 during the flare of SGR 1806-20.
 Based on the results of this latter work, our calculations suggest that low frequency QPOs, pictured by triangles, can be identified as dipole toroidal and dipole spheroidal
 nodeless vibration, respectively: $\nu(_0t_1)=18$ Hz and
 $\nu(_0s_1)=26$ Hz.}
 \end{figure}

 In Fig.3 set out very clearly the case for identifying two non-identified before
 low-frequency QPOs in data for SGR 1806-20 (Israel et al
 2005), namely $\nu=18$ Hz and $\nu=26$ Hz, highlighted by triangles.

\section{Summary}

 The obtained spectral formulae for the frequency of nodeless elastic vibrations trapped in the finite-depth
 seismogenic layer may be of some interest in its own right from the viewpoint of general theoretical
 seismology\cite{LW-95} in the sense that
 they can be utilized in the study of seismic vibrations of more wide class of solid celestial objects such
 as Earth-like planets.
 One of the remarkable feature of considered model is that the dipole nodeless overtones
 possess properties of Goldstone soft modes, that is, the dipole overtones emerge if and on only if
 the elastic vibrations turn out to be locked in the peripheral layer of finite
 thickness.
 It is shown that obtained spectral equations
 are consistent with the existence treatment of low-frequency QPOs in the X-ray luminosity of flares
 SGR 1900+14 and  SGR 1806-20 as caused by quake-induced torsional nodeless vibrations.
 What is newly disclosed here is that previously non-identified low-frequency QPOs in data for SGR 1806-20
 can be attributed to nodeless dipole torsional and spheroidal
 vibrations, namely, $\nu(_0t_1)=18\,\mbox{Hz}$ and $\nu(_0s_1)=26\,\mbox{Hz}$.

\section{Acknowledgements}

 This work is partly supported by NSC of Taiwan,
 under grants NSC-096-2811-M-007-012, NSC-096-2628-M-007-012-MY3 and NSC-097-2811-M-007-03.

\end{document}